\documentclass[12pt,a4paper,reqno]{amsart}
\usepackage{amsmath}
\usepackage{amsfonts}
\usepackage{amssymb}
\usepackage{graphicx,psfrag}                    
\usepackage{listings}                           
\usepackage{algorithm}
\usepackage{algorithmic}
\numberwithin{equation}{section}

\theoremstyle{plain}

\theoremstyle{definition}

\renewcommand{\geq}{\geqslant}

\newsavebox{\proofbox}
\savebox{\proofbox}{\begin{picture}(7,7)%
  \put(0,0){\framebox(7,7){}}\end{picture}}

\def\emph#1{{\it #1}}
\def\textbf#1{{\bf #1}}

\begin{document}

\title{The Period of the Subtraction Games}

\author{Zhihui Qin}

\email{zhihuiqindoris@gmail.com}

\author{Guanglei He}
\address{Sichuan University jinjiang college,CHINA
}
\email{guangleihe@gmail.com}

\begin{abstract}
Subtraction games is a class of impartial combinatorial games, They with finite subtraction sets are known to have periodic nim-sequences. So people try to find the regular of the games. But for specific of Sprague-Grundy Theory, it is too difficult to find, they obtained some conclusions just by simple observing. This paper used PTFN algorithm to analyze the period of the Subtraction games. It is more suitable than  \emph{Sprague-Grundy Theory}, and  this paper obtained four conclusions by PTFN algorithm . This algorithm provide a new direction to study the subtraction games' period.
\end{abstract}


\maketitle

\section{Introduction}

R. K. Guy maintains an ‘Unsolved Problems in Combinatorial Game Theory’ column.In the first column.

\emph{A1(1) Subtraction games with finite subtraction sets are known to have periodic nim-sequences. Investigate the relationship between the subtraction set and the length and structure of the period.}

A lot of papers(\cite{IJ1},\cite{JB},\cite{EJR},\cite{TS}) discussed this problem and observed the period of the game. Nonetheless,it is difficult to discover a generic mathematical expression,for the particularity of \emph{Sprague-Grundy Theory}. The important result published in \cite{RJN}.

For all $0 < s1_{1} <s2_{2}$ the $(s_{1},s_{2})$-game has period :

\begin{equation}
P =
\begin{cases}
2s_{1}& \text{if $ k = 3n , n \in{} N^{+}$ }\\
s_{1} + s_{2}& \text{if $ k \neq 3n , n \in{} N^{+}$}
\end{cases}
\end{equation}

In the \cite{GLH}, it put forward a new algorithm for subtraction games, it named PTFN algorithm. This paper used PTFN algorithm to analyse the period of the subtraction games. This paper feels that these features deserve explanation even though they occasionally fail.

\section{The Periodic of Subtraction Games}

PTFN algorithm makes us easy to find the subtraction games how to engender period .
Let us now consider to use the PTFN algorithm to analyse the period of the games.

For a $S = \{1,3,7,8\}$-games:
\begin{figure}[htb]
\centering
\includegraphics[width=4.2in]{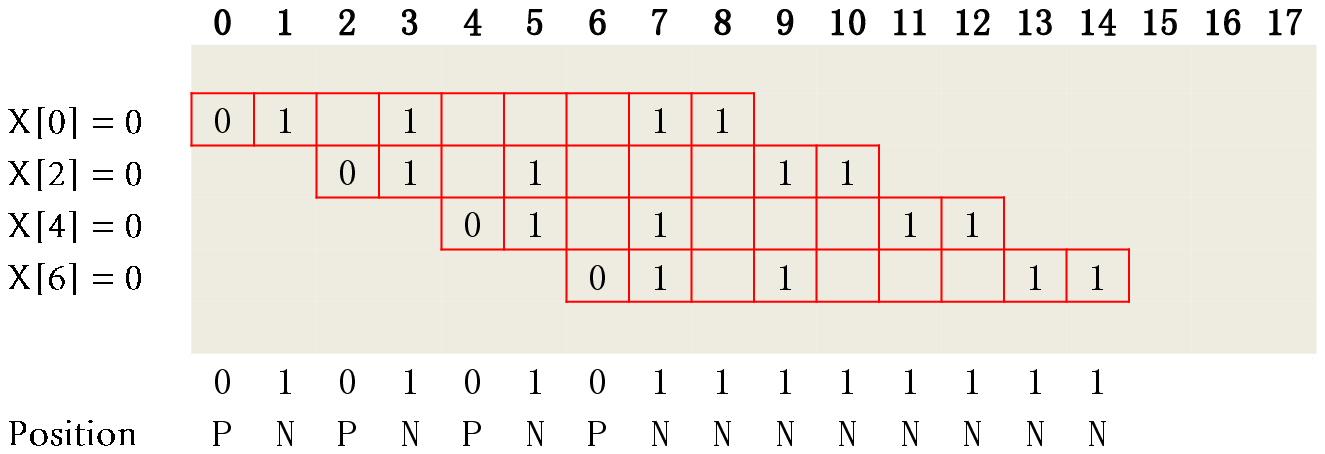}
\caption{How to get period}
\end{figure}

Base on the information contained in the Figure.1, this paper realized as the algorithm runs to $X[15]$ , all of the elements after $X[15]$ without computing , so $X[15]$ equal to $X[0]$ was also a terminal position. The result after $X[15]$ as same as $X[0]$. So it produce a cycle, the period is 15.

Our most prominent observation is :

\subsection*{Theorem 1:}

For all $k \geq 3$ the $(1,2,k)$-games have period :

\begin{equation}
P =
\begin{cases}
$k + 1$& \text{if $ k = 3n , n \in{} N^{+}$ }\\
3& \text{if $ k \neq 3n , n \in{} N^{+}$}
\end{cases}
\end{equation}

The result obtained is from table.1.
\begin{table}[htp]
\centering
\caption{Subtract Game with $s = \{1,2,k\}$}
\begin{tabular}{p{1.7cm}|p{6.7cm}| c}
S & Linear period & Period Length \\
\hline
1,2,3 & \.{0}11\.{1} & 4            \\
1,2,4 & \.{0}1\.{1} & 3             \\
1,2,5 & \.{0}1\.{1} & 3             \\
1,2,6 & \.{0}11011\.{1} & 7         \\
1,2,7 & \.{0}1\.{1} & 3             \\
1,2,8 & \.{0}1\.{1} & 3             \\
1,2,9 & \.{0}11011011\.{1} & 10     \\
1,2,10 & \.{0}1\.{1} & 3            \\
1,2,11 & \.{0}1\.{1} & 3            \\
1,2,12 & \.{0}11011011011\.{1} & 13 \\
1,2,13 & \.{0}1\.{1} & 3            \\
1,2,14 & \.{0}1\.{1} & 3            \\
1,2,15 & \.{0}11011011011011\.{1} & 16 \\
...         &...                           &... \\
\hline
\end{tabular}
\end{table}

\subsection*{Theorem 2:}

For all $k \geq 3$ the $(1,3,k)$-games have period :

\begin{equation}
P =
\begin{cases}
$k + 3$& \text{if $k$ is odd }\\
2& \text{if $k$ is even}
\end{cases}
\end{equation}

The result obtained is from table.2.

From the results obtained so far, it seems that the result of $S = \{1,2\}$ can cover all numbers which can't divided by 3. So the period of $S = \{1,2,k\}$-games is 3, k is the number can't divided by 3.
\begin{table}[htp]
\centering
\caption{Subtract Game with $s = \{1,3,k\}$}
\begin{tabular}{p{1.7cm}|p{6.7cm}| c}
S & Linear period & Period Length \\
\hline
1,3,4 & \.{0}10111\.{1} & 7                                     \\
1,3,5 & \.{0}\.{1} & 2 \\
1,3,6 & \.{0}1010111\.{1} & 9                                   \\
1,3,7 & \.{0}\.{1} & 2 \\
1,3,8 & \.{0}101010111\.{1} & 11                                \\
1,3,9 & \.{0}\.{1} & 2 \\
1,3,10 & \.{0}10101010111\.{1} & 13                             \\
1,3,11 & \.{0}\.{1} & 2 \\
1,3,12 & \.{0}1010101010111\.{1} & 15                           \\
1,3,13 & \.{0}\.{1} & 2 \\
1,3,14 & \.{0}101010101010111\.{1} & 17                         \\
1,3,15 & \.{0}\.{1} & 2 \\
1,3,16 & \.{0}10101010101010111\.{1} & 19                       \\
1,3,17 & \.{0}\.{1} & 2 \\
1,3,18 & \.{0}1010101010101010111\.{1} & 21                     \\
...         &...                   &...                         \\
\hline
\end{tabular}
\end{table}

\subsection*{Theorem 3:}
For all $k \geq 3$ the $(1,k,k + 1)$-games have period :
\begin{equation}
P =
\begin{cases}
$2k$& \text{if $k$ is odd }\\
2k + 1& \text{if $k$ is even}
\end{cases}
\end{equation}

The result obtained is from table.3.
\begin{table}[htp]
\centering
\caption{Subtract Game with $s = \{1,k,k + 1\}$}
\begin{tabular}{p{1.7cm}|p{6.7cm}| c}
S & Linear period & Period Length \\
\hline
1,2,3   &\.{0}11\.{1} & 4 \\
1,3,4   &\.{0}10111\.{1} & 7 \\
1,4,5   &\.{0}101111\.{1} & 8 \\
1,5,6   &\.{0}101011111\.{1} & 11 \\
1,6,7   &\.{0}1010111111\.{1} & 12 \\
1,7,8   &\.{0}1010101111111\.{1} & 15 \\
1,8,9   &\.{0}10101011111111\.{1} & 16 \\
1,9,10  &\.{0}10101010111111111\.{1} & 19 \\
1,10,11 &\.{0}101010101111111111\.{1} & 20 \\
1,11,12 &\.{0}101010101011111111111\.{1} & 23 \\
1,12,13 &\.{0}1010101010111111111111\.{1} & 24 \\
1,13,14 &\.{0}1010101010101111111111111\.{1} & 27 \\
1,14,15 &\.{0}10101010101011111111111111\.{1} & 28 \\
...         &...                   &...                         \\
\hline
\end{tabular}
\end{table}

\subsection*{Theorem 4:}
For $k_{i} = 0$ or $1$,The $(s,k_{1}(2s + 1),k_{2}(3s + 2),k_{3}(4s + 3),...,k_{n}((n + 1)s + n))$-game have period:
\begin{equation}
P = 2s
\end{equation}
\subsection*{A Sample proof } Based on the information(\textbf{Theorem 1 and 2} table.1 and table.2) before analyses, this paper knows that a set as consist of three elements' period has two situations; one is related to the set of the first two numbers , the other is related to the third number. So it has two results.

For \textbf{Theorem 3}, when $k$ is odd ,there isn't number be covered , so the period is $k + (k + 1)- 1 = 2k$. When $k$ is even, there is $k + 1$ be covered , so the period is $k + (k + 1) = 2k + 1$.

For \textbf{Theorem 4}, when $(s_{i} + 1)$ divided by $(s + 1)$,all others positions were covered by the first number .This assumption illustrated in (Figure.2):

\begin{figure}[htp]
\centering
\includegraphics[width=4.2in]{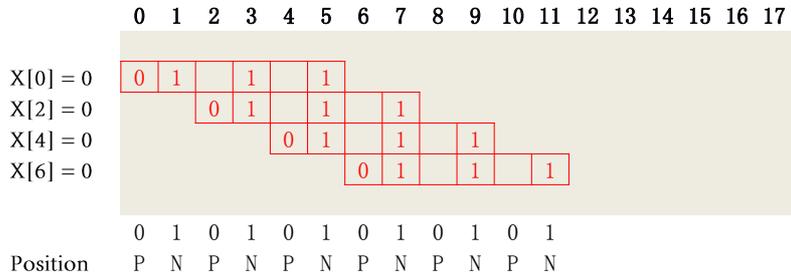}
\caption{Be covered}
\end{figure}
Therefore, the period of \textbf{Theorem 4} is equal to $S = \{s\}$ that is $2s$.

\section{Conclusions}
This paper used PTFN algorithm to analyse the period of subtraction games, finally it gets three assumptions(\textbf{Theorem 1,2,3}),Unfortunately, the study don't end for the period of subtraction games played are not obvious in some special cases (in Table.4).
\begin{table}[htp]
\centering
\caption{Subtract Game with special case}
\begin{tabular}{l|p{8cm}| c}
S & Linear period & Period\\
\hline
1,4,10      &010110101111101101\.{0}110110110\.{1}                              &11    \\
1,4,15      &010110101101011111011\.{0}10110101101101\.{1}                      &16    \\
1,4,20      &0101101011010110101111101101\.{0}1101011010110110110\.{1}          &21    \\
1,6,9       &010101101111\.{0}101\.{1}01011                                     &5     \\
1,6,14      &010101101010111111101101010110101101111\.{0}101\.{1}01011          &5     \\
1,6,16      &01010110101011011110101101\.{0}101\.{1}01011                       &5     \\
...         &...                                                                &...   \\
\hline
\end{tabular}
\end{table}

As shown in (in Table.4) for $S = \{1,4,5n\}$ $n \in N^{+}$ games,it has period even if it appeared late.Truly some subtraction games which the period is not obvious ，but has potential regular. How to find that need our deeper study.

\providecommand{\href}[2]{#2}

     \end{document}